\documentclass[prb,aps,twocolumn,eqsecnum,groupedaddress]{revtex4-2}
\usepackage[usenames,dvipsnames]{color}
\usepackage[colorlinks=true,linkcolor=Blue,citecolor=Blue,urlcolor=Blue]{hyperref}
\usepackage{graphicx}
\usepackage{bbm}
\usepackage{bm}
\usepackage{amsmath}
\usepackage{amssymb}
\usepackage{mathptmx}
\usepackage[version=3]{mhchem}

\newcommand{\ba}{\begin{eqnarray}}
\newcommand{\ea}{\end{eqnarray}}
\newcommand{\bd}{\begin{displaymath}}

\newcommand{\nn}{\nonumber \\}

\begin{document}
\title{Dipolar BF theory and dipolar braiding statistics}
%\author{Hyojae \surname{Jeon}}
%\affiliation{Department of Physics, Sungkyunkwan University, Suwon 16419, South Korea}
\author{Jung Hoon \surname{Han}}
\affiliation{Department of Physics, Sungkyunkwan University, Suwon 16419, South Korea}
\email{hanjemme@gmail.com}

\begin{abstract} We analyze the recently proposed dipolar BF theory with couplings to charge and dipole currents. The quasiparticles of the theory are either charge-like or dipole-like, and the mutual braiding statistics between charge-like and dipole-like quasiparticles are dipolar, meaning that it depends on the position of the quasiparticle being encircled. The braiding statistics between two dipole-like quasiparticles is that of ordinary anyons. We further prove that the dipolar BF theory is equivalent to the rank-2 tensor BF theory developed earlier as an effective theory for the rank-2 toric code. Although the two theories are equivalent, the dipolar BF formulation embodies the dipole symmetry explicitly and gives a clean insight into the way the dipole symmetry manifests itself in various conservation laws and the dipolar braiding statistics.  
\end{abstract}

\date{\today}
\maketitle

\section{Introduction} 
Extending the horizon of existing many-body models by incorporating dipolar, multipolar, or general modulated symmetries has been a fruitful theme of condensed matter theory in recent years. It was driven in part by a desire to incorporate fractonic behavior~\cite{nandkishore-fracton-review,pretko-fracton-review} into field-theoretic formulations~\cite{pretko18,seiberg-shao21,gorantla22}, and was further fueled by the experimental realization of tilted optical lattice and some interesting phenomena found in such systems~\cite{bakr20,aidelsburger21,weitenberg22,sala20,feldmeier20,sala22,lake22,lake23,feldmeier23,han23}. Now the converging view is that these extensions can be systematically categorized by the order of the multipole symmetry~\cite{gromov19,nandkishore21,gliozzi23,gromov24} being incorporated in the theory. Examples of extensions include rank-2 gauge theory coupled to fractonic quasiparticles~\cite{pretko18,prem18,seiberg-shao21}, Hubbard models with pair-hopping fermions or bosons~\cite{lake22,lake23,feldmeier23}, Chern-Simons theory with dipolar symmetry~\cite{prem18,you20,you21,ohmori23}, and an 1D SPT model protected by multipolar symmetry~\cite{dipolar-SPT}, to name a few.

Following an unrelated line of reasoning, we have written down a variant of the toric code called the rank-2 toric code (R2TC)~\cite{R2TC} by building on some earlier works~\cite{bulmash18,ma18}. Anyonic excitations in the R2TC have mobility restrictions not unlike the fractons and are forced to hop only by $N$ sites at a time in a $\mathbb{Z}_N$ stabilizer model of R2TC~\cite{R2TC,pace-wen}. Several other stabilizer models with similar mobility restrictions on anyonic excitations were discovered~\cite{delfino23,watanabe23,delfino23b}. The braiding statistics among the anyons in R2TC displayed unusual position dependence, distinct from the ordinary anyon braiding statistics that are purely topological. Attempts to explain the position-dependent braiding in the field theory framework were made by several authors~\cite{pace-wen,oh22-dBF}. In particular, the dipolar BF theory proposed in \cite{oh22-dBF} had rank-2 tensor gauge fields coupled to rank-2 tensor matter currents, and embodied dipole symmetry in its symmetry structure although the R2TC model had no such symmetry to begin with. In that regard the dipole symmetry appeared to be `emergent' rather than imprinted on the lattice Hamiltonian of R2TC from the outset. 

Quite recently, topological Chern-Simons and BF theories incorporating multipolar symmetries explicitly at the level of the effective action have been put forward~\cite{huang2023chern,ebisu2310}. They are indeed lucid representations of topological systems embodying both charge and dipole symmetries and are claimed to be equivalent to certain rank-2 Chern-Simons or BF theories after suitable transformations~\cite{prem18,you20,you21}. Here we look into this issue further and prove that the dipolar BF theory put forward in \cite{ebisu2310} is in fact equivalent to the effective theory of the R2TC previously suggested in \cite{oh22-dBF}, which was also dubbed the dipolar BF theory. To avoid the confusion, here we dub the new action the dipolar BF (dBF) theory, and rename the rank-2 BF theory of \cite{oh22-dBF} as the rank-2 BF (R2BF) theory. The origin of dipolar braiding statistics, first discussed field-theoretically in \cite{pace-wen,oh22-dBF}, is given a more transparent interpretation in the framework of dBF theory thanks to the dipole symmetry being embedded in the action from the outset. 

In Sec. \ref{sec:dBF} we re-visit the dBF action~\cite{ebisu2310} and analyze it in greater depth than \cite{ebisu2310} with additional coupling to matter fields. A complete set of conservation laws related to charge and dipole symmetries are derived. In Sec. \ref{sec:equivalence} we show that the dBF action can be transformed into the R2BF action and prove their equivalence. Some aspect of this analysis has been done in \cite{ebisu2310}, but we perform a more complete analysis with the full set of gauge-matter coupling allowed by the theory. Then in Sec. \ref{sec:dipolar-braiding} we re-visit the issue of dipolar braiding statistics and re-derive the results on the basis of insights gained from the analysis of dBF action. The paper is summarized in Sec. \ref{sec:discussion}. 

\section{Dipolar BF theory}
\label{sec:dBF}

The purpose of this section is to introduce the dipolar BF theory and outline its basic properties. Some outstanding features of dBF theory is shared by the dipolar Chern-Simons (dCS) theory~\cite{huang2023chern}, which is simpler and discussed first as a way of illustrating the main principle. 

The dCS theory consists of two parts ${\cal L} = {\cal L}_{\rm dCS} + {\cal L}_{AJ}$ where
\begin{align}
{\cal L}_{{\rm dCS}} & = \frac{N}{4\pi} \varepsilon_{\lambda \mu\nu} A_\lambda^a \partial_\mu A_\nu^a  \nn 
{\cal L}_{AJ} & = - J_\mu A_\mu - J_\mu^a A_\mu^a  . 
\end{align}
For each spatial direction $a=x,y$ we have the gauge field $A^a_\mu$ described by the Chern-Simons (CS) action with level $N$. Two kinds of matter fields $(J_\mu , J^a_\mu )$ are introduced, corresponding to the charge and the dipole current associated with each dipole moment orientation $a$ and coupled to the gauge fields $A_\mu$ and $A^a_\mu$, respectively. The summation over the spacetime indices $\mu, \nu = t , x, y$ and the spatial index $a=x,y$ is implicit. 

The gauge fields $A_\mu, A^a_\mu$ are not totally independent. Their gauge transformations are linked by the dipole conservation, and follow~\cite{ohmori23,huang2023chern,ebisu2310}
\begin{align}
    A^\nu_\mu & \rightarrow  A^\nu_\mu + \partial_\mu \beta^\nu , \nn 
    A_\mu & \rightarrow A_\mu + \partial_\mu \alpha + \beta^\mu . 
\label{eq:A-transform}   
\end{align}
Here $\beta^\mu$ is nonzero only for the spatial components $\mu = x, y$. There is no temporal component in the dipolar gauge fields, i.e. $A^t_\mu = 0$. The dCS action given above is invariant under the transformation \eqref{eq:A-transform}. There can be no CS term for $A_\mu$ or a mixture of $A_\mu $ and $A^a_\mu$ since such terms fail to remain invariant under \eqref{eq:A-transform}. Although $A_\mu$ does not participate in the CS action, it can still couple to the charge current $J_\mu$ and have non-trivial effect. Throughout the paper we refer to quantities without the superscript $a$ as `charge' and those with the superscript as `dipole' or `dipolar'. 

Requiring the invariance of the dCS action under the transformation \eqref{eq:A-transform} gives the continuity equations 
\begin{align}
\partial_\mu J_\mu = 0 , ~~
    \partial_\mu J_\mu^x = J_x ,  ~~ 
    \partial_\mu J_\mu^y = J_y . 
    \label{eq:charge-dipole-current-relation} 
\end{align}
While the charge current satisfies the usual continuity equation, the dipole currents $J_\mu^x , J_\mu^y$ satisfy  modified continuity equations with source terms coming from the charge current. Physically, it means that when a charge moves in a certain direction it will violate the dipole moment conservation in that direction and must act as a source term in the dipole continuity equation. One can re-write \eqref{eq:charge-dipole-current-relation} in a more suggestive form
\begin{align}
   \partial_\mu [ J_\mu^x - (x-x_0 ) J_\mu ]  = \partial_\mu [ J_\mu^y - (y-y_0) J_\mu ] = 0  
   \label{eq:dipole-continuity-eqs} 
\end{align}
where $x_0, y_0$ are some constants. We will make heavy use of this form when we discuss dipolar braiding in Sec. \ref{sec:dipolar-braiding}. 

We move to examine the dipolar BF theory given by~\cite{ebisu2310}
\begin{align} {\cal L} = {\cal L}_{\rm dBF} + {\cal L}_{AJ} + {\cal L}_{bK} , \label{eq:dBF} \end{align} 
where 
\begin{align}
{\cal L}_{{\rm dBF}} & = \frac{N}{2\pi} \varepsilon_{\lambda \mu\nu} b_\lambda \left(\partial_\mu A_\nu + A^\mu_\nu \right) + \frac{N}{2\pi} \varepsilon_{\lambda \mu\nu} b_\lambda^a  \partial_\mu A^a_\nu \nn 
{\cal L}_{AJ} & = - J_\mu A_\mu - J_\mu^a A_\mu^a  \nn 
{\cal L}_{bK} & =    -b_\mu K_\mu - b_\mu^a K_\mu^a . 
\label{eq:dBF}
\end{align}
The last term ${\cal L}_{bK}$ was not included in \cite{ebisu2310} but is necessary for a complete analysis. The gauge transformation properties of the dBF action are as follows:
\begin{align}
    A^\nu_\mu & \rightarrow  A^\nu_\mu + \partial_\mu \beta^\nu  \nn 
    A_\mu & \rightarrow A_\mu + \partial_\mu \alpha + \beta^\mu  \nn 
  b_\mu & \rightarrow b_\mu + \partial_\mu \theta \nn 
b_\mu^a & \rightarrow b_\mu^a +\partial_\mu \phi^a - \theta \delta_{\mu a} . 
\label{eq:Ab-transform}   
\end{align}
It is again assumed that $\beta^t = 0 = A^t_\mu$. 

Requiring the invariance of the dBF action under the transformation \eqref{eq:Ab-transform} gives several continuity equations:
\begin{align}
\partial_\mu J_\mu = 0 , ~~ \partial_\mu K_\mu^a  & = 0 \nn 
\partial_\mu [ J_\mu^x - (x-x_0 ) J_\mu ]  = \partial_\mu [ J_\mu^y - (y-y_0) J_\mu ] & = 0  \nn     
\partial_\mu [ K_\mu + (x-x_0 ) K^x_\mu + (y-y_0 ) K^y_\mu ] & = 0  . 
\label{eq:J-and-K-current-cons} 
\end{align}
These equations play vital roles in deducing braiding statistics among charge and dipoles, as discussed in Sec.~\ref{sec:dipolar-braiding}. 

\section{equivalence of dipolar BF theory to rank-2 BF theory}
\label{sec:equivalence}

We show how to gradually morph the dBF action in \eqref{eq:dBF} into the R2BF theory, written in terms of rank-2 tensor gauge fields and rank-2 tensor currents. In the end, we find that dBF action is in complete agreement with the R2BF action constructed in \cite{oh22-dBF} as an effective field theory of R2TC. Some aspect of this transformation was worked out in \cite{ebisu2310} for the pure gauge part ${\cal L}_{\rm dBF}$ without the coupling terms ${\cal L}_{AJ} + {\cal L}_{bK}$. 

The proof proceeds by first integrating out $b_t$ from the dBF action, resulting in the constraint:
\begin{align} A_{xy} - A_{yx} = (2\pi/N) K_t . \label{eq:Axy-Ayx} \end{align} 
We define the rank-2 gauge fields $A_{ab}$,
\begin{align} A_{ab} =\partial_a A_b - A_a^b , \label{eq:Aab} \end{align} 
which transforms under \eqref{eq:Ab-transform} as 
\begin{align} A_{ab} \rightarrow A_{ab} + \partial_a \partial_b \alpha . \end{align} 
Only the charge sector shows up in the transformation of $A_{ab}$. This is the well-known transformation associated with the tensor gauge fields previously used in fractonic field theories~\cite{pretko18,prem18,you20,you21}. The new insight here is that the tensor fields $A_{ab}$ are not the fundamental objects themselves, but are constructed from the underlying vector gauge fields $A_\mu^a$ and $A_\mu$. Although both $A_{xy} , A_{yx}$ transform in the same way, they are not the same fields when $K_t \neq 0$. The definition of $A_{ab}$ in \eqref{eq:Aab} does not guarantee such symmetry either. We will proceed for the time being by treating $A_{xy}$ and $A_{yx}$ as two distinct fields. 

Among the remaining terms in the action ${\cal L}$ (after integrating out $b_t$), terms containing $A^a_t$ can be collected into
\begin{align} \frac{N}{2\pi} \left[ A_t^a (\partial_x b_y^a - \partial_y b_x^a ) - b_y A^x_t + b_x A^y_t \right]  - A^a_t J_t^a . \label{eq:A_t^a} \end{align} 
Integrating out $A_t^a$ from \eqref{eq:A_t^a} enforces two relations:
\begin{align} 
    \partial_x b_y^x - \partial_y b_x^x - b_y & = (2\pi/N) J_t^x \nn 
    \partial_x b_y^y - \partial_y b_x^y + b_x & = (2\pi/N) J_t^y . 
    \label{eq:constraint-2} 
\end{align}

Since $b_t$ has been integrated out already, $\varepsilon_{\lambda \mu\nu} b_\lambda \partial_\mu A_\nu$ becomes $b_x (\partial_y A_t - \partial_t A_y ) + b_y (\partial_t A_x - \partial_x A_t ) $. We can use \eqref{eq:constraint-2} to write $b_x , b_y$ in this action in terms of other fields:
\begin{align}
& \frac{N}{2\pi} \Bigl[ b_x (\partial_y A_t - \partial_t A_y ) + b_y (\partial_t A_x - \partial_x A_t ) \Bigr]  \rightarrow  \nn 
    & \frac{N}{2\pi} \Bigl[ b_x^x (\partial_y \partial_t A_x - \partial_x \partial_y A_t ) - b_y^y (\partial_x \partial_t A_y - \partial_x \partial_y A_t ) \nn 
    & ~~~~~~~~ - b_x^y ( \partial_y^2 A_t -\partial_y \partial_t A_y ) + b_y^x (\partial_x^2 A_t - \partial_x \partial_t A_x ) \Bigr] \nn 
    & ~~~~~~~~~~~~ -A_t (\partial_x J_t^x + \partial_y J_t^y ) + A_x \partial_t J_t^x + A_y \partial_t J_t^y . 
    \label{eq:x}
\end{align}
In the meantime, the coupling $- J^a_x A_x^a - J^a_y A_y^a $ is re-written, using $A_{ab} = \partial_a A_b - A_a^b$, as 
\begin{align}
- ( J^a_x A_x^a + J^a_y A_y^a ) \rightarrow & A_x (\partial_x J_x^x + \partial_y J_y^x ) + A_y (\partial_x J^y_x + \partial_y J^y_y ) \nn 
    & + A_{xx} J_x^x + A_{yy} J_y^y + A_{xy} J_x^y + A_{yx} J_y^x .  
    \label{eq:yy} 
\end{align}
(Note that $A_t^a$ is already integrated out and no longer present in the action.) Collecting terms proportional to $A_\mu$ from the last line of \eqref{eq:x} and the first line of \eqref{eq:yy} and combining them with $-A_\mu J_\mu$, we conclude that $A_\mu J_\mu$ is modified to $A_\mu \tilde{J}_\mu$ where
\begin{align}
    \tilde{J}_t & = J_t + \partial_x J^x_t + \partial_y J^y_t \nn 
    \tilde{J}_x & = J_x - \partial_\mu J_\mu^x = 0  \nn 
    \tilde{J}_y & = J_y - \partial_\mu  J_\mu^y = 0 . 
\end{align}
The last two lines come from invoking the dipole current conservation \eqref{eq:charge-dipole-current-relation}. We thus arrive at the very simple result that only the temporal coupling $-A_t \tilde{J}_t$ survives in the charge sector of ${\cal L}_{AJ}$ after integrating out some fields. 
%\begin{align} & J_t^x (\partial_x A_t - \partial_t A_x ) + J_t^y (\partial_y A_t - \partial_t A_y )  \nn 
%& = -A_t (\partial_x J_t^x + \partial_y J_t^y ) + A_x \partial_t J_t^x + A_y \partial_t J_t^y . 
%\label{eq:y} 
%\end{align}
In the end we find
\begin{align} {\cal L}_{AJ} \rightarrow  -A_t \tilde{J}_t + A_{xx} J_{xx} +A_{xy} J_{xy} + A_{yx} J_{yx} + A_{yy} J_{yy} \label{eq:AJ-modified}
\end{align}
with 
\begin{align} 
\tilde{J}_t & = J_t + \partial_x J^x_t + \partial_y J^y_t \nn 
J_{xx} & = J_x^x, ~ J_{yy} = J_y^y, ~ J_{xy} = J_x^y, ~ J_{yx} = J_y^x .
\end{align} 

Other terms in the dBF action can be re-arranged as well, for instance 
\begin{align} & b_t^a (\partial_x A_y^a - \partial_y A_x^a ) \rightarrow \nn 
& ~~~~ b_t^x (-\partial_x A_{yx} + \partial_y A_{xx} ) + b_t^y (\partial_y A_{xy} - \partial_x A_{yy} ) \label{eq:xx}
\end{align}
using 
\begin{align} \partial_x A_y^x - \partial_y A_x^x  & = -\partial_x A_{yx} + \partial_y A_{xx } , \nn 
\partial_x A_y^y - \partial_y A_x^y & = -\partial_x A_{yy} + \partial_y A_{xy}.
\end{align} 
The other terms, $b_y^a \partial_t A_x^a - b_x^a \partial_t A_y^a $, can be combined with \eqref{eq:x} to give
\begin{align}
    & b_x^x (\partial_t A_{yx} - \partial_x \partial_y A_t )  - b_y^y (\partial_t A_{xy} - \partial_x \partial_y A_t ) \nn 
    & + b_x^y (\partial_t A_{yy} - \partial_y^2 A_t ) - b_y^x (\partial_t A_{xx} -\partial_x^2 A_t  ) . 
    \label{eq:xxx}
\end{align}
Finally, adding \eqref{eq:xx} and \eqref{eq:xxx} gives
\begin{align}
    {\cal L}_{\rm dBF} & = \frac{N}{2\pi} \Bigl[ b_t^x ( \partial_y A_{xx} - \partial_x A_{yx} ) + b_t^y (\partial_y A_{xy} - \partial_x A_{yy} ) \nn 
    & ~~~~~~ + b_x^x (\partial_t A_{yx} - \partial_x \partial_y A_t )  - b_y^y (\partial_t A_{xy} - \partial_x \partial_y A_t ) \nn 
    & ~~~~~~ + b_x^y (\partial_t A_{yy} - \partial_y^2 A_t ) - b_y^x (\partial_t A_{xx} - \partial_x^2 A_t ) \Bigr] . 
    \label{eq:dBF'1} 
\end{align}
In this form, $b_\mu$ has been completely integrated out and some second derivatives of the tensor gauge fields $A_{ab}$ appear in the action. 

So far we have avoided the assumption of symmetry $A_{xy} = A_{yx}$ and the action \eqref{eq:dBF'1} is valid even for $A_{xy} \neq A_{yx}$. To proceed further, we solve the constraint \eqref{eq:Axy-Ayx} by re-writing 
\begin{align}
    A_{xy} \rightarrow A_{xy} + \frac{\pi}{N} K_t ,
~~    A_{yx} \rightarrow A_{yx} - \frac{\pi}{N} K_t . 
    \label{eq:Axy-sym-antisym}
\end{align}
The new tensor is symmetric, $A_{xy} = A_{yx}$. We can re-write the action \eqref{eq:dBF'1} in light of the substitution \eqref{eq:Axy-sym-antisym}. In addition to the second line of \eqref{eq:dBF'1} simplifying to $(b_x^x - b_y^y) (\partial_t A_{xy} - \partial_x \partial_y A_t )$, there are additional terms showing up in the action due to $K_t$: 
\begin{align}
-\frac{1}{2} b_x^x \partial_t K_t  
-\frac{1}{2} b_y^y \partial_t K_t  
+\frac{1}{2} b_t^x \partial_x K_t 
+\frac{1}{2} b_t^y \partial_y K_t  . 
\label{eq:32}
\end{align}
Also invoking \eqref{eq:constraint-2}, we can transform the $bK$ coupling 
\begin{align} b_x K_x + b_y K_y \rightarrow  b_y^y \partial_x K_x + b_x^x \partial_y K_y - b_x^y \partial_y K_x - b_y^x \partial_x K_y . \label{eq:33} \end{align} 
We can add \eqref{eq:32} and \eqref{eq:33} to $b_\mu^a K_\mu^a$ to get
\begin{align}
&     b_t^x \left( K_t^x -\frac{1}{2} \partial_x K_t \right) + b_t^y \left( K_t^y -\frac{1}{2} \partial_y K_t  \right) \nn 
& + b_x^x \left(K_x^x  + \partial_y K_y +\frac{1}{2} \partial_t K_t \right) + b_y^x \left(K_y^x -\partial_x K_y \right) \nn 
& + b_x^y \left(K_x^y -\partial_y K_x \right) + b_y^y \left(K_y^y + \partial_x K_x + \frac{1}{2} \partial_t K_t \right) \nn
& \equiv b^a_\mu \tilde{K}_\mu^a , 
\end{align}
and conclude the action ${\cal L}_{bK}$ is modified to
\begin{align} {\cal L}_{bK} \rightarrow - b^a_\mu \tilde{K}_\mu^a . 
\label{eq:bK-modified} 
\end{align}
The charge part $-b_\mu K_\mu$ has been integrated out. Meanwhile, the ${\cal L}_{AJ}$ action in \eqref{eq:AJ-modified} is modified by $K_t \neq 0$, 
\begin{align} {\cal L}_{AJ} \rightarrow  -A_t \tilde{J}_t + A_{xx} J_{xx} +A_{xy} J_{xy} + A_{yy} J_{yy} + \frac{\pi}{N} (J_x^y - J_y^x ) K_t ,
\label{eq:AJ-modified-2} 
\end{align}
with $A_{xy} = A_{yx}$ and $J_{xy} = J_x^y + J_y^x$. 

It turns out $K_t$ appears linearly in \eqref{eq:bK-modified} and \eqref{eq:AJ-modified-2} but does not appear elsewhere. Integrate out $K_t$ gives
\begin{align}
( \partial_y b_t^y  - \partial_t b_y^y ) - (\partial_t b_x^x - \partial_x b_t^x ) = \frac{2\pi}{N} (J_x^y - J_y^x ) . 
\end{align}
This relation is nothing new, and can be derived from the constraint that arises from integrating out $A^a_x , A^a_y$ from the full action:
\begin{align} 
\partial_y b_t^a - \partial_t b_y^a & = (2\pi/N) J^a_x \nn 
\partial_t b_x^a - \partial_x b_t^a & = (2\pi/N) J^a_y . 
\end{align}

Finally, we can arrange the two actions ${\cal L}_{AJ}$ and ${\cal L}_{bK}$ as
\begin{align}
    {\cal L}_{AJ} & = -A_t \tilde{J}_t + A_{xx} J_{xx} +A_{xy} J_{xy} + A_{yy} J_{yy} \nn
    {\cal L}_{bK} & = -b_\mu^a \tilde{K}_\mu^a 
    \label{eq:AJ-bK-final}
\end{align}
with
\begin{align}
& \tilde{J}_t = J_t + \partial_x J^x_t + \partial_y J^y_t , \nn 
& J_{xx} = J_x^x , ~~ J_{xy} = J_x^y + J_y^x , ~~ J_{yy} = J_y^y ,   
\label{eq:J-defined}
\end{align}
and
\begin{align} \tilde{K}_t^a & = K_t^a \nn 
\tilde{K}_x^x & = K_x^x  + \partial_y K_y , ~~~ \tilde{K}_y^x = K_y^x -\partial_x K_y \nn 
\tilde{K}_x^y & = K_x^y -\partial_y K_x  , ~~~ \tilde{K}_y^y = K_y^y + \partial_x K_x . 
\end{align} 
We note that ${\cal L}_{\rm dBF}$ and ${\cal L}_{AJ}$ are by now written in terms of tensorial gauge fields and currents, but ${\cal L}_{bK}$ is not. 

To remedy this problem, we re-define certain variables 
\begin{align} (b_t^x , b_t^y ) & \rightarrow (E_t^y , - E_t^x ), \nn 
(b_x^x - b_y^y , b_x^y , b_y^x ) & \rightarrow (- E_{xy} , - E_{yy} , E_{xx} ), 
\label{eq:b-to-E} \end{align} 
%
%${\cal L}'_{{\rm dBF}, 1}$ becomes 
%
%\begin{align}
%{\cal L}'_{{\rm dBF}, 1} & =  A_{xy} \partial_t E_{xy}  + A_{yy} \partial_t E_{yy} + A_{xx} \partial_t E_{xx}  \nn & +  E_t^x (\partial_x A_{yy} - \partial_y A_{xy}) + E_t^y ( \partial_y A_{xx} - \partial_x A_{xy} ) \nn 
%& +  A_t (\partial_x \partial_y E_{xy} + \partial_y^2 E_{yy} + \partial_x^2 E_{xx} ) . 
%\end{align} 
and further re-write
\begin{align} (E_{xx}, E_{yy}, E_{xy} ) & \rightarrow (E_{yy}, E_{xx}, -E_{xy}) , \nn  
(A_{xx}, A_{yy}, A_{xy} ) & \rightarrow (A_{yy}, A_{xx}, -A_{xy}) . 
\label{eq:rewrite-A-and-E} 
\end{align} 
This gives
\begin{align}
{\cal L}_{\rm dBF} & \rightarrow A_{xy} \partial_t E_{xy}  + A_{yy} \partial_t E_{yy} + A_{xx} \partial_t E_{xx} \nn 
& ~~ + E_t^x ( \partial_x A_{xx} + \partial_y A_{xy}  ) + E_t^y ( \partial_y A_{yy} + \partial_x A_{xy} )  \nn 
& ~~ +  A_t (\partial_y^2 E_{xx} + \partial_x^2 E_{yy} - \partial_x \partial_y E_{xy}) , 
\end{align} 
and 
\begin{align}{\cal L}_{AJ} = -A_t \tilde{J}_t + A_{yy} J_{xx} + A_{xx} J_{yy} - A_{xy} J_{xy} .
\end{align}
Meanwhile, re-writing ${\cal L}_{bK}$ with these new variables gives 
\begin{align}
{\cal L}_{bK} \rightarrow  & - E_t^y \tilde{K}_t^x +E_t^x \tilde{K}_t^y + E_{xx} \tilde{K}_x^y - E_{yy} \tilde{K}_y^x - \frac{1}{2} E_{xy} (\tilde{K}_x^x - \tilde{K}_y^y ) \nn
& - \frac{1}{2} (b_x^x + b_y^y ) (\tilde{K}_x^x + \tilde{K}_y^y ) 
\label{eq:L_bK'} 
\end{align}
The last line cannot be written in terms of any of the variables in \eqref{eq:b-to-E}. Luckily, $b_x^x + b_y^y$ does not appear elsewhere in the action and can be integrated out to yield
\begin{align} \tilde{K}_x^x + \tilde{K}_y^y = K_x^x + K_y^y + \partial_\mu K_\mu = 0, 
\label{eq:K-constraint} 
\end{align} 
which is nothing but the $K$-current conservation already derived in \eqref{eq:J-and-K-current-cons}. 

Combining all three remaining actions, the full action ${\cal L} ={\cal L}_{\rm dBF} + {\cal L}_{AJ} + {\cal L}_{bK}$ becomes 
\begin{align}
{\cal L}_{\rm R2BF} = & \frac{N}{2\pi} \Bigl[ A_{xy} \partial_t E_{xy}  + A_{yy} \partial_t E_{yy} + A_{xx} \partial_t E_{xx} \nn 
& ~~~ ~~~ + E_t^x ( \partial_x A_{xx} + \partial_y A_{xy} -\rho^x ) \nn 
& ~~~ ~~~ + E_t^y ( \partial_y A_{yy} + \partial_x A_{xy} - \rho^y )  \nn 
& +  A_t (\partial_y^2 E_{xx} + \partial_x^2 E_{yy} - \partial_x \partial_y E_{xy} - \tilde{J}_t ) \Bigr] \nn 
& ~~~ ~~~ -A_{xy} J_{xy} + A_{yy} J_{xx} + A_{xx} J_{yy} \nn 
& ~~~ ~~~ - E_{xx} K_{xx} - E_{xy} K_{xy} - E_{yy} K_{yy} ,
\label{eq:R2BF} 
\end{align} 
with
\begin{align}
(\rho^x , \rho^y ) & = (- \tilde{K}_t^y ,  \tilde{K}_t^x) \nn 
(K_{xx} , K_{xy} , K_{yy} ) & = \left( -\tilde{K}_x^y , \frac{1}{2} (\tilde{K}_x^x - \tilde{K}_y^y )  ,  \tilde{K}_y^x \right) . 
\label{eq:rho-K-defined} 
\end{align}
The final action ${\cal L}_{\rm R2BF}$ is in complete accord with the R2BF action derived in \cite{oh22-dBF} [Eq. (5.2)]. The R2BF action supports a scalar charge $\tilde{J}_t$ and the associated tensor current $(J_{xx}, J_{xy}, J_{yy})$, as well as vector charges $( \rho^x , \rho^y )$ and the associated tensor current $(K_{xx}, K_{xy}, K_{xy})$. Although there are two vector charges, only one tensor current $K_{ab}$ exists for both of them.  

There are three conservation laws, one for each of the three charges. One of them is given by $(\tilde{J}_t , J_{xx}, J_{xy}, J_{yy})$ satisfying the equation
\begin{align} & \partial_t \tilde{J}_t + \partial_x^2 J_{xx} + \partial_x \partial_y J_{xy} + \partial_y^2 J_{yy} \nn 
& = \partial_t J_t + \partial_x (\partial_\mu J^x_\mu ) + \partial_y (\partial_\mu J^y_\mu ) \nn 
& = \partial_t J_t + \partial_x J_x + \partial_y J_y = 0 . 
\label{eq:cons-1}
\end{align} 
The relation $\partial_\mu J^a_\mu = J_a$ was used to reach the second line. The charge density $\tilde{J}_t$ is the sum of $J_t$, the bare charge density, and the dipole-induced charge density $\partial_x J^x_t + \partial_y J^y_t$. A charge ($J_t$) by itself cannot move due to the dipole moment conservation and must be accompanied by dipole motion. This forces charge to form a composite object with a dipole in order to move freely. Although the observation that a free object in dipole-constrained theory is a charge-dipole composite has been made in \cite{oh22-dBF} already, the dBF formulation is helpful in clarifying the nature of this composite charge as the sum of the bare charge ($J_t$) and the dipole-induced charge ($\partial_x J^x_t + \partial_y J^y_t$) through the formula \eqref{eq:J-defined}. The tensor currents $(J_{xx} , J_{xy} , J_{yy} )$ and $(K_{xx} , K_{xy} , K_{yy} )$, introduced through heuristic reasoning in \cite{oh22-dBF}, are also given concrete definitions in terms of underlying dipole currents. 

The second continuity equation follows from defining a three-current $$(\rho^x, K_{xx} , K_{xy} ) = (-\tilde{K}^y_t , -\tilde{K}^y_x , (\tilde{K}^{xx}-\tilde{K}^{yy})/2 )$$ and working out
\begin{align}
    & \partial_t \rho^x + \partial_x K_{xx} + \partial_y K_{xy}  \nn 
    & = \partial_t \rho^x + \partial_x K_{xx} + \partial_y K_{xy} - \frac{1}{2} \partial_y (\tilde{K}_x^{x} +  \tilde{K}_y^{y} ) \nn 
    & = - \partial_\mu \tilde{K}_\mu^y = - \partial_\mu K^y_\mu =  0 .
    \label{eq:cons-2}
\end{align}
Note that in the second line we added an extra term $\partial_y ( \tilde{K}_x^{x} + \tilde{K}_y^{y} )/2$ to complete the continuity equation. This is allowed since the added term is zero by the constraint 
\eqref{eq:K-constraint}. It also implies a degree of flexibility in defining $K_{xy}$, as
\begin{align} K_{xy} = \frac{1}{2} (\tilde{K}_x^x - \tilde{K}_y^y ) + \frac{\theta}{2} (\tilde{K}_x^x + \tilde{K}_y^y ) \label{eq:Kxy} \end{align}
with $\theta$ being an arbitrary real number. 

A third conservation law exists for the three-current $$(\rho^y , K_{xy}, K_{yy} ) = (\tilde{K}^x_t , (\tilde{K}_x^{x}-\tilde{K}_y^{y})/2  , \tilde{K}^x_y )$$ obeying the continuity equation
\begin{align} 
    & \partial_t \rho^y + \partial_x K_{xy} + \partial_y K_{yy}  \nn 
    & = \partial_t \rho^y + \partial_x K_{xy} + \partial_y K_{yy} + \frac{1}{2} \partial_x (\tilde{K}_x^{x} +  \tilde{K}_y^{y} ) \nn 
    & = \partial_\mu \tilde{K}_\mu^x = \partial_\mu K^x_\mu =  0 .  
\label{eq:cons-3} 
\end{align}

In conclusion, we have derived all three conservation laws as well as the exact form of the R2BF given in \cite{oh22-dBF} by starting from the dBF action of \cite{oh22-dBF} and integrating out some degrees of freedom. In the course of the transformation we found explicit representations of the various charge densities and tensor currents in the R2BF in terms of the charge and dipole currents of the original dBF - see 
\eqref{eq:J-defined} and \eqref{eq:rho-K-defined}. These expressions will play a vital role in deriving dipolar braiding statistics. 

\section{Dipolar Braiding}
\label{sec:dipolar-braiding}

A key feature of the dBF (or R2BF) theory is that its quasiparticles show braiding statistics that depends on the location of the particle being braided around~\cite{R2TC,oh22-dBF}. This point can be highlighted already in the dCS theory, which we discuss first due to its simplicity. 

It is well known that $\partial_\mu J_\mu =0$ admits a solution 
\begin{align} J^c_\mu = (1, \dot{x}_c , \dot{y}_c ) \delta^2 ({\bf r} - {\bf r}_c (t) ) 
\label{eq:single-particle-solution}
\end{align} 
where ${\bf r}_c (t)$ is the position of the charge at time $t$, and $(\dot{x}_c , \dot{y}_c )$ is its velocity. 
By virtue of continuity equation $\partial_\mu [J_\mu^x - (x-x_0 ) J_\mu ] = 0$ and $\partial_\mu [J_\mu^y - (y -y_0 ) J_\mu ] = 0$ [Eq. \eqref{eq:dipole-continuity-eqs}], one has solutions for the dipole currents 
\begin{align} J_\mu^x & = (x_c -x_0 ) J_\mu^{c} + J_\mu^{d,x} \nn 
J_\mu^y & = (y_c -y_0 ) J_\mu^{c} + J_\mu^{d,y} 
\label{eq:induced-Jx-Jy} \end{align}
with $J_\mu^c$ given in \eqref{eq:single-particle-solution} and 
\begin{align} J^{d,x}_\mu & = (1, \dot{x}_{x} , \dot{y}_{x} ) \delta^2 ({\bf r} - {\bf r}_{x} (t) ) \nn 
J^{d,y}_\mu & = (1, \dot{x}_{y} , \dot{y}_{y} ) \delta^2 ({\bf r} - {\bf r}_{y} (t) )
\label{eq:single-xy-dipole-solution}
\end{align} 
representing currents associated with a single $x$- or $y$-dipole located at ${\bf r}_{x}(t)$ and ${\bf r}_{y}(t)$, respectively. We will refer to solutions given in \eqref{eq:single-particle-solution} and \eqref{eq:single-xy-dipole-solution} as single-charge and single-dipole solutions, respectively. Current due to a single-charge motion $J_\mu^c$ generates accompanying dipole current $J^x_\mu = (x-x_0) J^c_\mu$ and $J^y_\mu = (y-y_0) J^c_\mu$. To distinguish them from the {\it intrinsic} dipole currents $J^{d,x}_\mu$ and $J^{d,y}_\mu$, we refer to them as {\it orbital} dipole currents of $a$-orientation. The name was suggeested in \cite{huang2023chern}.

We examine the consequence of inserting various solutions of the charge and the dipole currents into the action ${\cal L}_{AJ}$. First, inserting the single-charge solution $J_\mu = J^c_\mu$ into $J_\mu A_\mu$ (and integrating over spacetime) gives
\begin{align}
\oint [ A_x ({\bf r}_c ) dx_c + A_y ({\bf r}_c ) dy_c ] = \int (\partial_x A_y - \partial_y A_x ) dx_c dy_c  
\label{eq:charge-part-of-phase} 
\end{align}
where the integral is over the area encompassed by the closed trajectory of ${\bf r}_c$. Since there is no Chern-Simons term for $A_\mu$, this flux is not quantized. In fact, $\partial_x A_y - \partial_y A_x$ is not even gauge-invariant under the dipolar gauge transformation by $\beta$ in \eqref{eq:A-transform}, and will soon be combined with another term to make it fully gauge-invariant. 

When we insert the single-dipole solution \eqref{eq:single-xy-dipole-solution} into the action $J_\mu^a A^a_\mu$, we find for $a=x$ 
\begin{align} \int J_\mu^{d,x} A_\mu^x = \int (\partial_x A^x_y - \partial_y A^x_x ) . \label{eq:Jmux} \end{align} 
Invoking the flux attachment rule 
\begin{align} \varepsilon_{\lambda \mu\nu} \partial_\mu A^a_\nu = \frac{2\pi}{N} J^a_\lambda , \label{eq:dipole-flux-attachment} \end{align} 
which can be derived by integrating out $A^a_\mu$ from the dCS action, one can further write
\begin{align} \int (\partial_x A^x_y - \partial_y A^x_x ) =\frac{2\pi}{N} \int J_t^x 
\end{align} 
showing that the geometric phase is $2\pi/N$ per $x$-dipole enclosed within the path. We conclude that an $a$-dipole braiding around another dipole of the same orientation results in the fractional phase of $2\pi/N$. 

When a charge is being braided, there is an accompanying orbital dipole current $(x-x_0) J_\mu^c$ contributing to $J^x_\mu$. Its contribution to the phase is   
\begin{align}
& \oint (x-x_0 ) J_\mu^c A_\mu^x= \int (x_c -x_0) [ A_x^x ({\bf r}_c ) dx_c + A_y^x ({\bf r}_c ) dy_c ] \nn 
%& = \int \left( \partial_x [(x_c -x_0 ) A^x_y] - \partial_y [(x_c -x_0 )A^x_x ] \right) dx_c dy_c \nn 
& = \oint  (x - x_0 )[ \partial_x A_y^x - \partial_y A_x^x ] dx dy + \int A_y^x dx dy . 
\label{eq:x-dipole-part-of-phase} 
\end{align}
There is also the orbital $y$-dipole current $(y-y_0) J_\mu^c$ contributing the phase 
\begin{align}
& \oint (y-y_0 ) J_\mu^c A_\mu^y = 
%\int (x_c -x_0) [ A_x^x ({\bf r}_c ) dx_c + A_y^x ({\bf r}_c ) dy_c ] \nn 
%& = \int \left( \partial_x [(x_c -x_0 ) A^x_y] - \partial_y [(x_c -x_0 )A^x_x ] \right) dx_c dy_c \nn 
\oint  (y - y_0 )[ \partial_x A_y^y - \partial_y A_x^y ] dx dy - \int A_x^y dx dy . 
\label{eq:y-dipole-part-of-phase}
\end{align}
Overall, there are three different contributions to the phase, given in \eqref{eq:charge-part-of-phase}, \eqref{eq:x-dipole-part-of-phase}, \eqref{eq:y-dipole-part-of-phase}, as a charge is moved adiabatically around. Combining all three contributions gives
\begin{align}
& \oint  (x - x_0 )[ \partial_x A_y^x - \partial_y A_x^x ] dx dy  +    \oint  (y - y_0 )[ \partial_x A_y^y - \partial_y A_x^y ] dx dy \nn 
& + \int (\partial_x A_y - \partial_y A_x + A_y^x - A_x^y ) dx dy  . 
\end{align}
The second line represents a contribution from the gauge-invariant flux~\footnote{The author thanks Xiaoyang Huang for pointing out the gauge invariance of this combination.} $F_{xy} = \partial_x A_y - \partial_y A_x + A_y^x - A_x^y$, but the value is not tied to the charge density as there is no CS term involving $A_\mu$ in the dCS action. The first line, after substituting $\partial_x A_y^a -\partial_y A_x^a = (2\pi/N)J_t^a$, gives
\begin{align} - \frac{2\pi}{N} \Bigl[  \sum_i (x_i - x_0 ) + \sum_i (y_i - y_0 ) \Bigr] \label{eq:x-dipole-phase} \end{align} 
where $x_i$ ($y_i$) is the $a$-coordinate of the $i$-th $a$-dipole inside the loop. The minus sign occurs because the action is ${\cal L}_{AJ} = - A_\mu J_\mu - A_\mu^a J_\mu^a$. The braiding statistics between a charge and the dipoles are therefore {\it dipolar}, in the sense that the phase acquired in the braiding process depends linearly on the coordinates of the particles.

In summary, when an $a$-dipole is adiabatically moved, it picks up the fractional phase of $2\pi/N$ by braiding around other $a$-dipoles. Meanwhile a charge braiding around dipoles picks up dipolar phase because the charge motion induces orbital dipole current. The orbital dipole current accompanying the charge current is necessitated by the need to satisfy the dipole symmetry. 

Next we examine the braiding statistics in the dBF (or R2BF) theory. From the various conservation laws of $J$-currents and $K$-currents derived in \eqref{eq:J-and-K-current-cons}, we can write down the solutions 
\begin{align} & J_\mu = J^c_\mu  \nn 
& J^x_\mu = J^{d,x}_\mu + (x -x_0 ) J^c_\mu , ~~ J^y_\mu = J^{d,y}_\mu + (y -y_0 ) J^c_\mu  \nn 
& K^a_\mu = K^{d,a}_\mu  \nn 
& K_\mu = K^{c}_\mu - (x -x_0 ) K^{d,x}_\mu - (y -y_0 ) K^{d,y}_\mu , 
\label{eq:various-constraints} \end{align}
where $J^c_\mu , K^c_\mu , J^{d,a}_\mu , K^{d,a}_\mu$ are the single-particle solutions of \eqref{eq:single-particle-solution} and \eqref{eq:single-xy-dipole-solution}. 

Since our focus is on the dipolar braiding statistics, we can ``turn off" the intrinsic dipolar current $J^{d,a}_\mu$ and write the orbital dipole currents as 
\begin{align} 
J^x_\mu & = (x-x_0 ) J^c_\mu = (x - x_0 ) (1, \dot{x}_c , \dot{y}_c ) \delta^2 ( {\bf r} - {\bf r}_c ) \nn 
J^y_\mu & = (y-x_0 ) J^c_\mu = (y - x_0 ) (1, \dot{x}_c , \dot{y}_c ) \delta^2 ( {\bf r} - {\bf r}_c ) . 
\end{align} 
This gives
\begin{align}
    \tilde{J}_t & = J_t + \partial_x J^x_t + \partial_y J^y_t \nn 
    & = [1 + (x - x_0 ) \partial_x + (y - y_0 ) \partial_y ]  \delta^2 ( {\bf r} - {\bf r}_c )  \nn 
    J_{xx} & = J_x^x = (x - x_0 ) \dot{x}_c \delta^2 ( {\bf r} - {\bf r}_c ) \nn 
    J_{xy} & = J_x^y + J_y^x = [ (x - x_0 ) \dot{y}_c + (y - y_0 ) \dot{x}_c ] \delta^2 ( {\bf r} - {\bf r}_c ) \nn 
    J_{yy} & = J_y^y = (y - y_0 ) \dot{y}_c \delta^2 ( {\bf r} - {\bf r}_c )  . 
\end{align}
Inserting $(J_{xx}, J_{xy} , J_{yy})$ obtained here into the action $\int (A_{xx} J_{yy} + A_{yy} J_{xx} - A_{xy} J_{xy} )$ gives
\begin{align} \oint {\bf a} ({\bf r}_c ) \cdot d{\bf r}_c 
\end{align} 
where
\begin{align}
{\bf a}  = \Bigl( (x-x_0 ) A_{yy} - (y- y_0 ) A_{xy} , (y-y_0 ) A_{xx} - (x-x_0 ) A_{xy} \Bigr) .  \nonumber 
\end{align}
Taking the curl of ${\bf a}$ gives
\begin{align} {\bm \nabla} \times {\bf a} & = (y-y_0 ) (\partial_x A_{xx} + \partial_y A_{xy} ) - (x-x_0 ) (\partial_x A_{xy} + \partial_y A_{yy} ) \nn 
& = \frac{2\pi}{N} \left[ (y-y_0 ) \rho^x - (x-x_0 ) \rho^y \right] ,
\end{align} 
where we used the relations $\partial_x A_{xx} + \partial_y A_{xy} = \rho^x$ and $\partial_y A_{yy} + \partial_x A_{xy} = \rho^y$ obtained from integrating out $E_t^x, E_t^y$ in the R2BF action. Since we further have $\rho^x = - K^y_t$ and $\rho^y  = K_t^x$, we can re-write the curl 
\begin{align} {\bm \nabla} \times {\bf a} = -  \frac{2\pi}{N} \left[ (x-x_0 ) K_t^x + (y-y_0 ) K_t^y  \right] . 
\end{align} 
The Berry phase acquired by moving a $J$-charge along a closed path is therefore
\begin{align} - \frac{2\pi}{N} \left[  \sum_i (x_i -x_0 ) + \sum_i (y_i - y_0 ) \right] , \end{align}
where $x_i$ ($y_i$) are the $x$- ($y$-) coordinate of the $x$-dipole ($y$-dipole) enclosed inside the path. 
Adiabatic motion of $J$-charge is interpreted as that of a bare $J$-charge and $J$-dipole composite. 

We next turn to single-dipole solutions for $\partial_\mu K_\mu^a = 0$:
\begin{align}
    K_\mu^{d,x} & = (1, \dot{x}_{x} , \dot{y}_{x} ) \delta^2 ({\bf r} - {\bf r}_{x} ) \nn 
    K_\mu^{d,y} & = (1, \dot{x}_{y} , \dot{y}_{y} ) \delta^2 ({\bf r} - {\bf r}_{y} ) . 
\end{align}
The $K$-dipoles, being exempt from dipole constraint, can move freely. The solutions for $\partial_\mu [ K_\mu + (x- x_0 ) K_\mu^x + (y- y_0 ) K_\mu^y ] = 0$ is
\begin{align} 
K_\mu = - (x - x_0 ) K_\mu^{d,x} - (y - y_0 ) K_\mu^{d,y} . 
\end{align}
It is the sum of contributions from $x$-dipoles and $y$-dipoles. One can keep only one of the terms on the right  when there are dipoles of one orientation present but not the other. 

The two vector charges $\rho^x = -K^y_t = -\delta^2 ({\bf r} -{\bf r}_{y} )$ and $\rho^y = K^x_t = \delta^2 ({\bf r} -{\bf r}_{x} )$ refer to situations where there is only a single $y$-dipole or a single $x$-dipole. When we work out the associated $K$-currents $(K_{xx}, K_{xy} , K_{yy})$, we can separately consider situations with only the $x$-dipole or the $y$-dipole present. For instance, we choose $\rho^x = -\delta^2 ({\bf r} -{\bf r}_{y} )$ and the associated $K$-currents are
\begin{align}
K_{xx} = - \tilde{K}^y_x & \rightarrow  -K_x^{d,y} - (y - y_0 ) \partial_y K_x^{d,y} , \nn 
K_{xy} = \tilde{K}_x^x  & \rightarrow  -(y - y_0 ) \partial_y K_y^{d,y} , \nn 
K_{yy} = \tilde{K}_y^x & \rightarrow (y - y_0 ) \partial_x K_y^{d,y} . 
\label{eq:K-for-y-dipole} 
\end{align}
That is, from the generic definition of $(\tilde{K}^y_x , \tilde{K}_x^x , \tilde{K}_y^x )$ given in \eqref{eq:rho-K-defined}, we extract those single-dipole solutions associated with the $y$-dipole (terms on the right side of the arrow). 

Inserting \eqref{eq:K-for-y-dipole} into the R2BF action gives
\begin{align}
 -\int (E_{xx} K_{xx} + E_{xy} K_{xy} + E_{yy} K_{yy} ) \rightarrow  \oint {\bf a}^y \cdot d{\bf r}_y 
\end{align}
where
\begin{align} {\bf a}^y = \Bigl( E_{xx} - (y - y_0 ) \partial_y E_{xx} , (y - y_0 ) (\partial_x E_{yy} - \partial_y E_{xy} ) \Bigr) . \end{align} 
Taking the curl of ${\bf a}^y$ gives the dipolar flux attachment rule:
\begin{align} {\bm \nabla} \times {\bf a}^y & = (y-y_0 ) (\partial_x^2 E_{yy} + \partial_y^2 E_{xx} - \partial_x \partial_y E_{xy} ) \nn 
& = \frac{2\pi}{N} (y -y_0 ) \tilde{J}_t . \label{eq:curl-of-ay} \end{align} 
The second line follows from integrating out $A_t$ in the R2BF action which gives $\partial_x^2 E_{yy} + \partial_y^2 E_{xx} - \partial_x \partial_y E_{xy} = (2\pi/N) \tilde{J}_t$. The phase picked up by a $y$-dipole is 
\begin{align} \frac{2\pi}{N} \sum_i (y_i - y_0 ) \end{align}
from braiding around the composite $J$-particles enclosed inside, in addition to the fractional phase $2\pi/N$ acquired by braiding around other $y$-dipoles. 

A similar conclusion is reached for the braiding of $x$-dipole. For $\rho^y = \delta^2 ({\bf r} -{\bf r}_{x} )$, we find
\begin{align}
K_{xx} = \tilde{K}^y_x& \rightarrow  - (x - x_0 ) \partial_y K_x^{d,x} , \nn 
K_{xy} = -\tilde{K}_y^y  & \rightarrow  (x - x_0 ) \partial_x K_x^{d,x} , \nn 
K_{yy} = \tilde{K}_y^x & \rightarrow  K_y^{d,x} + (x - x_0 ) \partial_x K_y^{d,x}
\label{eq:K-for-x-dipole}
\end{align}
as the $K$-current associated with the adiabatic motion of $x$-dipole. Inserting \eqref{eq:K-for-x-dipole} into the R2BF action gives
\begin{align}
-\int (E_{xx} K_{xx} + E_{xy} K_{xy} + E_{yy} K_{yy} ) \rightarrow \oint {\bf a}^x \cdot d{\bf r}_x
\end{align}
where
\begin{align} {\bf a}^x = \Bigl( (x - x_0 ) ( \partial_x E_{xy} - \partial_y E_{xx}),(x -x_0 ) \partial_x E_{yy} - E_{yy} \Bigr) . \end{align} 
Taking the curl of ${\bf a}^x$ results in the dipolar flux attachment rule:
\begin{align} {\bm \nabla} \times {\bf a}^x = \frac{2\pi}{N} (x- x_0 ) \tilde{J}_t . \end{align} 
Braiding an $x$-dipole around the composite $J$-particles results in the dipolar phase
\begin{align} \frac{2\pi}{N} \sum_i (x_i - x_0 ) , \end{align} 
in addition to $2\pi/N$ from braiding around other $x$-dipoles. 

To conclude, dipolar braiding occurs between a composite $J$-charge and $K$-dipoles and are proportional to the $x$ ($y$) coordinate of the particle being braided around when the $K$-dipole involved is an $x$-dipole ($y$-dipole). The braiding statistics between $K$-dipoles of the same orientation is that of the usual anyons with fractional phases. 

% We write down 
% %
% \begin{align}
%     \rho^x & = -K^y_t = -\delta^2 ({\bf r} -{\bf r}_{d,x} ) \nn 
%     K_{xx} & = -\tilde{K}_x^y = -K_x^y +\partial_y K_x = - \dot{x}_{d,x} \delta^2 ({\bf r} -{\bf r}_{d,x} )  \nn 
%     &  - (x_{d,x}- x_0 ) \dot{x}_{d,x} \partial_y \delta^2 ({\bf r} -{\bf r}_{d,x} ) - (y_{d,y} - y_0 ) \dot{x}_{d,y} \partial_y \delta^2 ({\bf r} - {\bf r}_{d,y} )  \nn 
% %
% K_{xy} & = -\tilde{K}_x^x = -K_x^x - \partial_y K_y = - \dot{x}_{d,x} \delta^2 ({\bf r} - {\bf r}_{d,x} ) \nn 
% & +  (x_{d,x} - x_0 ) \dot{y}_{d,x} \partial_y \delta^2 ({\bf r} - {\bf r}_{d,x} ) +  (y_{d,y} - y_0 ) \partial_y \dot{y}_{d,y} \delta^2 ({\bf r} - {\bf r}_{d,y} ) \nn 
% %
%     K_{yy} & = \tilde{K}_y^x = K_y^x - \partial_x K_y = \dot{y}_{d,x} \delta^2 ({\bf r} -{\bf r}_{d,x} ) \nn 
%     & + (x_{d,x} - x_0 ) \dot{y}_{d,x} \delta^2 ({\bf r} - {\bf r}_{d,x} ) + (y_{d,y} - y_0 ) \dot{y}_{d,y} \delta^2 ({\bf r} - {\bf r}_{d,y} ) . 
% \end{align}

% Before feeding these solutions into the action, it is convenient to summarize various constraints relating the gauge fields to the currents:
% %
% \begin{align}
%     \varepsilon_{\lambda \mu \nu} (\partial_\mu A_\nu + A^\mu_\nu ) & = (2\pi/N) K_\lambda \nn 
%     \varepsilon_{\lambda \mu \nu} \partial_\mu A^a_\nu & = (2\pi/N) K^a_\lambda \nn 
%     \varepsilon_{\lambda \mu \nu} \partial_\mu b_\nu & = (2\pi/N) J_\lambda \nn
%     \partial_x b_y^x - \partial_y b_x^x - b_y  & = (2\pi/N) J_t^x \nn 
%     \partial_x b_y^y - \partial_y b_x^y + b_x  & = (2\pi/N) J_t^y . 
% \end{align}

\section{Discussion}
\label{sec:discussion}

The picture underlying the rank-2 toric code was that of two layers of $\mathbb{Z}_N$ lattice gauge theories being coupled through some constraint~\cite{R2TC,oh23}. The constraint was not related to the dipole symmetry in any way, or at least the connection to dipole symmetry was not obvious. The dipolar braiding statistics among the anyonic excitations in the rank-2 toric code was therefore an unexpected surprise, and its understanding prompted the formulation of BF-like theory employing rank-2 tensor gauge fields~\cite{oh22-dBF}. The dipolar braiding among the quasiparticles was derived from the rank-2 BF theory, but the precise relation between the dipolar braiding statistics and the dipole symmetry of the action remained obscure. In this work, the origin of dipolar braiding among the anyons is clarified as a direct consequence of the dipole symmetry built into the dipolar BF theory. 

Analysis of the present paper suggests that dipolar braiding is a natural consequence of topological models with dipole symmetry. Extension of the ideas and techniques explored in this paper to the BF theory with quadrupolar symmetry~\cite{ebisu2310} with the resulting quadrupolar braiding statistics should be straightforward, albeit technically challenging. 

\acknowledgments I am grateful to Xiaoyang Huang and H. Ebisu for an insightful exchange on the dipolar Chern-Simons and dipolar BF theories. Many discussions on the rank-2 toric code and dipolar braiding statistics with Jintae Kim, Yun-Tak Oh, and Yizhi You over the years are gratefully acknowledged. He was supported by the National Research Foundation of Korea(NRF) grant funded by the Korea government(MSIT) (No. 2023R1A2C1002644). This research was also supported in part by grants NSF PHY-1748958 and PHY-2309135 to the Kavli Institute for Theoretical Physics (KITP).

\bibliography{ref}
\end{document}